\documentclass[reprint,
 amsmath,amssymb,
 aps,
prstab,
floatfix,
]{revtex4-1}

\usepackage{graphicx}
\usepackage{dcolumn}
\usepackage{bm}
\usepackage{upgreek}

\begin{document}

\preprint{APS/123-QED}

\title{Obliquely Incident Laser and Electron Beam Interaction in an Undulator}
\thanks{Work supported by the National Nature Science Foundation of China (Grant No. 11605277) and National Key Research and Development Program of China (No. 2016YFA0401901).}%

\author{Xiaofan Wang \textsuperscript{1,2}}
\author{Chao Feng\textsuperscript{1,3}}
\email{fengchao@sinap.ac.cn}
\author{Cheng-Ying Tsai\textsuperscript{4}}
\author{Li Zeng\textsuperscript{1,2}}
\author{Zhentang Zhao\textsuperscript{1,3}}
\email{zhaozhentang@sinap.ac.cn}
\affiliation{
		\textsuperscript{1}Shanghai Institute of Applied Physics,  Chinese Academy of Sciences, Shanghai 201800, China \\
        \textsuperscript{2}University of Chinese
Academy of Sciences, Shijingshan District,
Beijing 100049, China\\
        \textsuperscript{3}Shanghai Advanced Research Institute, Chinese Academy of Sciences, Shanghai 201210, China\\       
        \textsuperscript{4}Huazhong University of Science and Technology, Wuhan 430074, China}


\begin{abstract}
The angular drift of a laser beam is of particular concern in applications such as seeded 
free-electron lasers. A systematical study of the obliquely incident laser and electron beam interaction in an undulator is presented in this paper. Theoretical analysis and numerical simulations demonstrate that the interaction would imprint energy and angular modulations on the electron beam simultaneously. Compared with the normally incident pattern, the obliquely incident laser-electron interaction leads to reductions in the bunching factors of HGHG and EEHG. In the meanwhile, proactive applications of this multi-dimensional modulation technique may bring vitality to the field of laser-electron manipulation.

\end{abstract}

\maketitle


\section{introduction}

X-ray free-electron lasers (FELs) based on electron linear accelerators (linacs) hold the promise for opening up new frontiers of ultra-fast and ultra-small sciences at the atomic length scale. This has been proved with the operation of x-ray FEL facilities such as Free-Electron laser in Hamburg (FLASH)~\cite{ackermann2007operation}, the Linac Coherent Light Source (LCLS)~\cite{emma2010first}, the Spring-8 Angstrom Compact free-electron Laser (SACLA)~\cite{ishikawa2012compact}, the Pohang Accelerator Laboratory X-ray Free Electron Laser (PAL-XFEL)~\cite{ko2017construction} and the Trieste FERMI~\cite{allaria2012highly}. Most of these operational facilities are based on the self-amplified spontaneous emission (SASE) mode \cite{kondratenko1980generation,bonifacio1984collective}, where the spontaneous radiation amplification process starts from electron beam shot noise. SASE FEL has the capability of providing extremely high-intensity, ultra-short light pulses with excellent spatial coherence but poor temporal coherence and relatively large shot-to-shot power fluctuations. The temporal coherence of the x-ray SASE FEL can be appreciably improved by the self-seeding scheme \cite{feldhaus1997possible,saldin2001x,geloni2011novel,amann2012demonstration}, which employs an x-ray monochromator sandwiched by double undulator configuration. The SASE radiation generated by the first undulator is purified by the monochromator and then further exponentially amplified to saturation in the second undulator.

An alternative way to significantly improve the temporal coherence and inhibit fluctuations of high-gain FELs is to use frequency up-conversion schemes, which generally rely on the techniques of optical-scale precise manipulation of the electron beam phase space with external coherent laser sources. The most famous frequency up-conversion scheme is the so-called high-gain harmonic generation (HGHG) \cite{yu1991generation}, which consists of two undulators separated by a dispersive section. A normally incident seed laser pulse is used to interact with electrons in the first undulator (modulator) to generate a sinusoidal energy modulation of the electron beam at the lasing wavelength. This energy modulation is then transformed into an associated spatial density modulation by a dispersion element. Coherent radiation at shorter wavelength is generated after the micro-bunched electron beam traverses the second undulator (radiator), which is tuned to a high harmonic of the seed frequency. Because the HGHG output is a direct map of the seed laser's attributes, this ensures a higher degree of temporal coherence and smaller pulse energy fluctuations with respect to SASE. 

However, significant bunching at higher harmonics usually requires enhancing energy modulation in HGHG, which would result in a degradation of the amplification in the radiation process. Thus, the excessive dependence on energy modulation prevents it from reaching higher harmonics through a single-stage HGHG. Cascaded HGHG \cite{yu1997high} is proposed to overcome this problem. It uses the FEL generated by the previous radiator as the seed laser for the following HGHG stage with the help of the "fresh bunch" technique. This scheme has been adopted by the FERMI and the Shanghai soft X-ray Free-Electron Laser facility (SXFEL) \cite{zhao2017status} as the main operation mode. Later, the echo-enabled harmonic generation (EEHG) \cite{stupakov2009using,xiang2009echo} emerges to improve the frequency multiplication efficiency in a single FEL amplification stage, which employs two modulators and two dispersive sections. This more complicated phase space manipulation technique partially shifts the dependence on energy modulation to the dispersion, thereby enabling high harmonic jumps with a relatively small laser-induced energy spread. The lasing of EEHG at EUV and soft x-ray wavelength regions have been achieved at SXFEL and FERMI recently~\cite{feng2019coherent,ribivc2019coherent}, paving the way for user experiments in the near future.

In these frequency up-conversion schemes, it is expected that the seed lasers should be coaxial superimposed (or should propagate coaxially) with the electron beam in the modulators. 
However, the laser angular drift caused by machine vibration or laser pointing instability will make the incident electromagnetic (EM) plane wave has a small angle relative to the propagation direction of the electron beam. This is very critical for the lasing of a seeded FEL at very high harmonics of the seed. It has been experimentally observed at SXFEL that under certain circumstances, for either HGHG or EEHG, the intensity of the FEL is very sensitive to the incident angle of the seed laser. Hereby a systematic study on the laser-electron interaction 
with a cross angle in the undulator is necessary for seeded FELs.

In this work, we firstly present the theoretical study and simulations of the obliquely incident laser and electron beam interaction in the undulator in Sec.~\ref{sec2}. Results show that the interaction would imprint not only energy modulation but also angular modulation on the electron beam phase space. Then in Sec.~\ref{sec3}, the dependences of the seed laser incident angle on the bunching factors of HGHG and EEHG are investigated, respectively. Finally, conclusions are given in Sec.~\ref{sec4}.

\section{Theoretical analysis and simulations}\label{sec2}
Figure 1
\begin{figure}
    \includegraphics[width=1\linewidth]{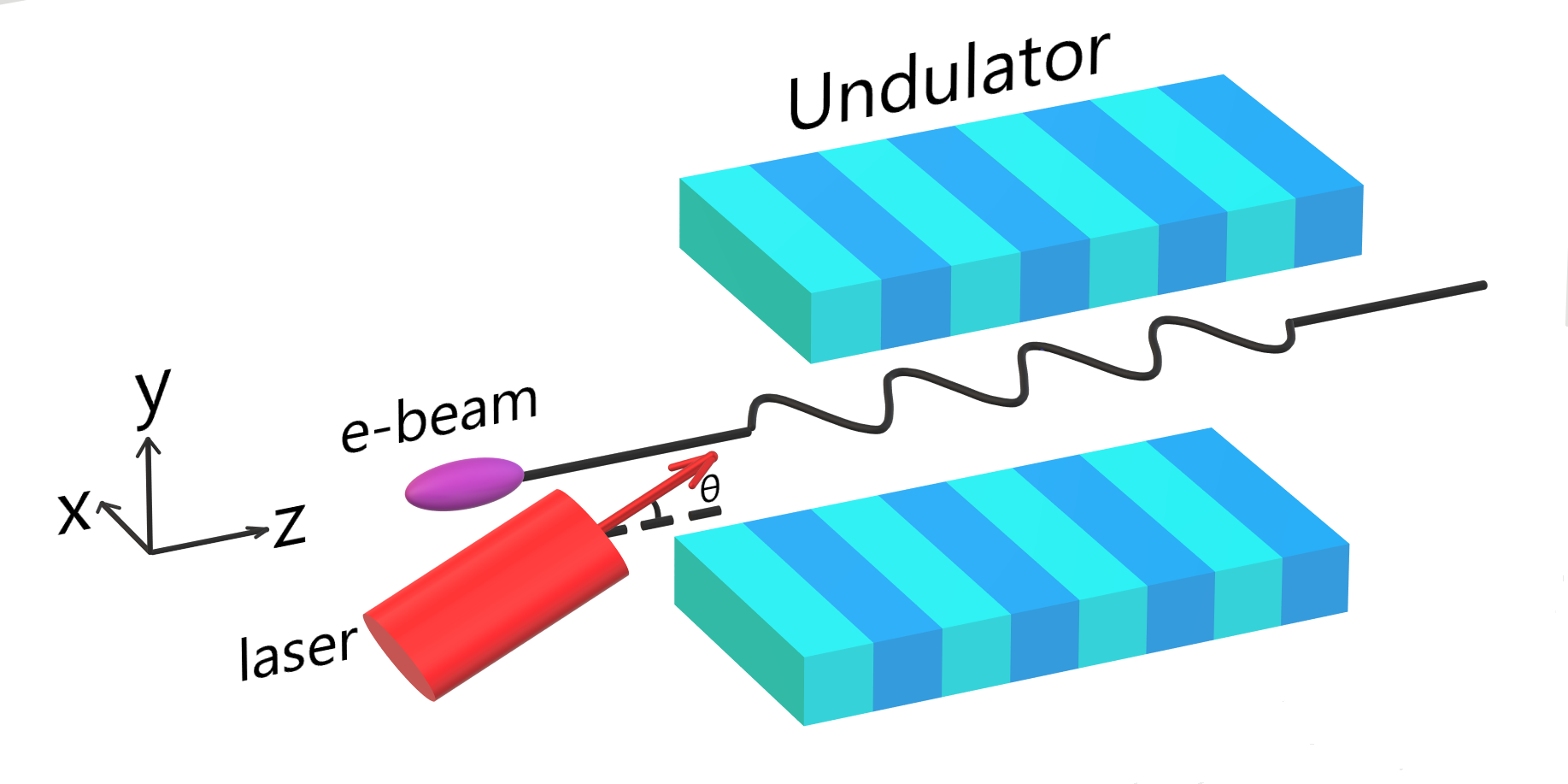}
    \caption{Schematic view of the sine-like trajectory of the electrons and of the oblique incidence with a small angle $\theta$ of the electromagnetic plane wave in a planar undulator. The laser field is polarized in the horizontal plane.}
    \label{fig1}
\end{figure}
schematically shows the laser-electron beam interaction with a cross angle, where one can find the sine-like trajectory of the electrons and the oblique incidence of a seed laser in a planar undulator. Cartesian coordinates are given here, where $x$, $y$ and $z$ represent the horizontal, vertical and longitudinal coordinates, respectively. The planar undulator has a sinusoidal magnetic field of period $\lambda_u$ and peak amplitude $B_u$:
\begin{eqnarray}\label{eq1}
B_y=B_u \cos{(k_u z)},
\end{eqnarray}
where $k_u = 2\pi/\lambda_u$. Under the influence of the Lorentz force generated by the undulator magnetic field, the normalized horizontal and longitudinal velocities of the electron beam are
\begin{subequations} 
\begin{eqnarray}
\beta_{x}&=&-\frac{K}{\gamma}\sin{(k_u z)},\\
\beta_{z}&\approx& 1-\frac{1}{2\gamma^2}(1+\frac{K^2}{2})+\frac{K^2}{4\gamma^2}\cos{(2k_u z)},
\end{eqnarray}
\end{subequations}
where $K=\frac{e B_u\lambda_u}{2\pi m c}=0.934B_u[\text{Tesla}]\lambda_u[\text{cm}]$ is the undulator parameter, $\gamma$ is the relativistic factor, $c$ is the speed of light, $e$ and $m$ are the electron charge and mass, respectively. Then the undulator-induced motions of the electrons are deduced:
\begin{subequations}
\begin{eqnarray}
x&=&\frac{K}{k_u\gamma}\cos{(k_u z)},\\
z&\approx& ct-\frac{1}{2\gamma^2}(1+\frac{K^2}{2})ct+\frac{K^2}{8k_u\gamma^2}\sin{(2k_u z)},
\end{eqnarray}
\end{subequations}
where $t$ is the electron traveling time in the undulator. The positions of the electrons are expressed as $\mathbf{x}=x_0+x$, $\mathbf{y}=y_0$, $\mathbf{z}=z_0+z$, where $x_0$ and $y_0$ are repectively the electron initial horizontal and vertical coordinates at the entrance of the undulator, $z_0$ is the electron coordinate within the electron bunch. 

The resonant condition occurs when the forward co-propagation radiation slips a distance $\lambda_s$ relative to the electrons after one undulator period. This principle is also applied to the cases where the EM wave is obliquely incident with a small angle: 
\begin{eqnarray}
\lambda_u \cos{\theta}+\lambda_s=\frac{\lambda_u}{\bar{\beta_{z}}},
\end{eqnarray}
where $\bar{\beta_{z}}$ is the average value of $\beta_{z}$. With $\cos{\theta}\approx 1-\frac{\theta^2}{2}$ and $\frac{1}{\bar{\beta_{z}}}\approx 1+\frac{1}{2\gamma^2}(1+\frac{K^2}{2})$, one finds the resonant expression for the laser-electron interaction with an oblique incidence in the undulator:
\begin{eqnarray}\label{eq5}
\lambda_s=\frac{\lambda_u}{2\gamma^2}(1+\frac{K^2}{2}+\gamma^2\theta^2).
\end{eqnarray}
Under this condition, the interaction between the electrons and the EM wave can be sustained, and a net transfer of energy from the electrons to the EM wave occurs. The resonance condition also shows that, for effective laser-beam interaction, the laser incidence angle should be on the order of $1/\gamma$ or smaller. 

The seed laser is an EM plane wave polarized in the horizontal plane. For simplicity, we assume that the transverse size of the EM wave is much larger than the electron beam, then one can neglect the transverse dependence of the field in the vicinity of the tightly collimated electron beam. The pulse length of the seed laser is assumed much longer than the electron bunch length, so we can neglect the laser power variation within the bunch. Considering a stationary coordinate system ($x_l$, $y_l$, $z_l$) where $x_l$ and $z_l$ respectively represents the polarization and pointing direction of the EM wave, the laser field can be simply written as
\begin{equation}\label{eq6}
E_{x_l}=E_0  \sin{\big(k_s(z_l-ct)\big)},
\end{equation}
where $E_0$ is the field amplitude, $k_s = 2\pi/\lambda_s$ is the wave number. Due to the spatially random uncertainty of laser pointing, the angle drift of an EM wave may occur in the $z - y$ plane, the $z - x$ plane, or somewhere between them. For simplicity, two cases are taken into consideration here: the laser incident in the $z - y$ plane and the laser incident in the $z - x$ plane.

\subsection{Laser incident in the $z - y$ plane}
When the laser is incident in the $z - y$ plane with a small angle relative to the $z$ coordinate, the geometric transformation from the laser stationary coordinate system to the Cartesian coordinate system reads
\begin{subequations}\label{eq7}
\begin{eqnarray}
x_l&=&\mathbf{x},\\
y_l&=&\mathbf{y}\cos{\theta}-\mathbf{z}\sin{\theta},\\
z_l&=&\mathbf{z}\cos{\theta}+\mathbf{y}\sin{\theta}.
\end{eqnarray}
\end{subequations}
Velocities follow the same transformation principle. Then the laser field can be written as
\begin{equation}\label{eq8}
E_{x}=E_0  \sin{\big(k_s(z\cos{\theta}-ct)+\Phi\big)},
\end{equation}
where $\Phi=k_s (z_0\cos{\theta}+y_0\sin{\theta})$. When the electron beam and the EM wave co-propagate through the undulator, they interact with each other. The instantaneous energy change of the electron beam is
\begin{eqnarray}\label{eq9}
\frac{\mathrm{d}\gamma}{\mathrm{d}z}&=&\frac{e}{m c^2}E_{x}\beta_{x}.
\end{eqnarray}
Integrating over the entire undulator length, the dimensionless energy deviation with respect to the reference particle is
\begin{eqnarray}
\Delta\delta&=&\frac{\Delta\gamma}{\gamma}=\frac{e E_0}{\gamma m c^2}\int_{0}^{N_u \lambda_u} \sin{\big(k_s(z\cos{\theta}-ct)+\Phi\big)}\cdot\beta_{x} dz\nonumber\\
&=&-\frac{e K E_0 N_u \lambda_u}{\gamma^2 m c^2} \Big\langle\sin{\big(k_s(z \cos{\theta}-ct)+\Phi\big)} \nonumber\\
&&\times\sin{(k_u z)}\Big\rangle_z.
\end{eqnarray}
Brackets denote the average of the z coordinate over one undulator period $\langle\cdots\rangle_z=\frac{1}{\lambda_u}\int_0^{\lambda_u} (\cdots) dz$. The argument in the first sine function can be expressed as
\begin{widetext}
\begin{eqnarray}
k_s(z\cos{\theta}-ct)&=&k_s(\cos{\theta}-1)ct-\frac{k_s}{2\gamma^2}(1+\frac{K^2}{2})ct\cos{\theta}+\frac{K^2 k_s}{8k_u\gamma^2}\sin{(2k_u z)}\cos{\theta}\nonumber\\
&\approx &-k_s\frac{\theta^2}{2}z-\frac{k_s}{2\gamma^2}(1+\frac{K^2}{2})z+\frac{K^2 k_s}{8k_u \gamma^2}\sin{(2k_u z)} = -k_u z+\frac{K^2/4}{1+K^2/2+\gamma^2\theta^2}\sin{(2k_u z)}\nonumber\\
&=&-k_u z+\frac{\xi}{2}\sin{(2k_u z)} ,
\end{eqnarray}
where we let $z\approx c t$ to obtain the first-order solution and $\xi=K^2/(2+K^2+2\gamma^2\theta^2)$. With the help of Bessel's integrals $J_{\alpha}(x)=\frac{1}{2\pi}\int_{0}^{2\pi}\cos{(\alpha \tau-x\sin{\tau})}\, \mathrm{d}\tau$, we find
\begin{eqnarray}
\Big\langle\sin{\big(k_s(z\cos{\theta}-ct)+\Phi\big)}\cdot\sin{(k_u z)}\Big\rangle_z&=&-\frac{1}{2}\Big\langle\cos{\big(\frac{\xi}{2}\sin{(2k_u z)}+\Phi\big)}-\cos{\big(-2k_u z+\frac{\xi}{2}\sin{(2k_u z)}+\Phi\big)}\Big\rangle_z
\nonumber\\
&=&-\frac{\cos{\Phi}}{2}\Big\langle\cos{\big(\frac{\xi}{2}\sin{(2k_u z)}\big)}-\cos{\big(-2k_u z+\frac{\xi}{2}\sin{(2k_u z)}\big)}\Big\rangle_z\nonumber\\
&=&-\frac{\cos{\Phi}}{2}(J_0(\frac{\xi}{2})-J_{1}(\frac{\xi}{2}))=-\frac{\cos{\Phi}}{2}\mathbf{J},
\end{eqnarray}
\end{widetext}
where $\mathbf{J}=J_0(\xi/2)-J_1(\xi/2)$, $J_0$ and $J_1$ are the first- and the second-order Bessel functions of the first kind. Then the energy modulation due to the laser-electron interaction can be written as
\begin{eqnarray}
\Delta\delta&=&\frac{e K E_0 N_u \lambda_u \mathbf{J}}{2\gamma^2 m c^2}\cos{\Phi}=\Lambda \cos{\Phi}\nonumber\\
&=&\Lambda \cos{(k_s (z_0\cos{\theta}+y_0\sin{\theta}))},
\end{eqnarray}
where $\Lambda=\frac{e K E_0 N_u \lambda_u \mathbf{J}}{2\gamma^2 m c^2}$. From this formula, one can see that the electrons may be accelerated or decelerated, depending on the relative phase of electron initial position.

Now we define the instantaneous change of electron vertical divergence due to the interaction with the laser electromagnetic field in the undulator:
\begin{eqnarray}
\frac{\mathrm{d}y^{\prime}}{\mathrm{d}z}&=&\frac{e}{\gamma m c^2}B_{y_l}\sin{\theta} v_{x}=\frac{e E_x}{\gamma m c^2}\sin{\theta} \beta_{x},
\end{eqnarray}
where $y^{\prime}=\frac{\mathrm d y}{\mathrm d z}=\beta_y$ is the normalized vertical velocity. Following the derivation procedure of the energy modulation, one obtains the angular modulation expression:
\begin{eqnarray}
\Delta y^{\prime}=\Lambda \sin{\theta}\cos{(k_s (z_0\cos{\theta}+y_0\sin{\theta}))}.
\end{eqnarray}
The energy and angular changes of the electron beam are in agreement with the Panofsky-Wentzel theorem~\cite{panofsky1956some}. Equation~(15) can be directly obtained from Eq.~(13) by using the theorem:
\begin{eqnarray}
   \frac{\partial}{\partial y_0} (\Delta\delta) = \frac{\partial \Delta y^{\prime}}{\partial z_0}.
\end{eqnarray}
Because the incident angle $\theta$ is typically small,  Eq.~(13) and Eq.~(15) can be simplified to
\begin{subequations}
\begin{eqnarray}
\Delta\delta&=&\Lambda\cos{(k_s z_0+k_s y_0\theta)},\\
\Delta y^{\prime}&=&\Lambda \theta\cos{(k_s z_0+k_s y_0\theta)}.
\end{eqnarray}
\end{subequations}
\subsection{Laser incident in the $z - x$ plane}
When the laser is incident in the $z - x$ plane with a small angle relative to the $z$ coordinate, the geometric transformation from the laser stationary coordinate system to the Cartesian coordinate system reads
\begin{subequations}
\begin{eqnarray}
x_l&=&\mathbf{x}\cos{\theta}-\mathbf{z}\sin{\theta},\\
y_l&=&\mathbf{y},\\
z_l&=&\mathbf{z}\cos{\theta}+\mathbf{x}\sin{\theta}.
\end{eqnarray}
\end{subequations}
Velocities follow the same transformation principle. Then the laser field can be written as
\begin{eqnarray}
E_{x_l}=E_0  \sin{\big(k_s(z\cos{\theta}+x\sin{\theta}-ct)+\Phi\big)},
\end{eqnarray}
where $\Phi=k_s (z_0\cos{\theta}+x_0\sin{\theta})$. Different from Eq.~(8), a term associated with the horizontal motion of the electron beam in the undulator appears in the expression of the electric field. When the electron beam and the EM wave co-propagate through the undulator, the instantaneous energy change of the electron beam now reads 
\begin{eqnarray}
\frac{\mathrm{d}\gamma}{\mathrm{d}z}&=&\frac{e}{m c^2}E_{x_l}\beta_{x_l}=\frac{e}{m c^2}E_{x_l}(\beta_{x} \cos{\theta}-\beta_{z} \sin{\theta}),
\end{eqnarray}
where components of the laser electric field in the $x$ and $z$ directions all contribute to the energy change. This indicates that the laser incident in the $z$ - $x$ plane is much more complicated than the vertical plane case. Integrating over the entire undulator length, the dimensionless energy deviation with respect to the reference particle is
\begin{eqnarray}
\Delta\delta&=&\frac{\Delta\gamma}{\gamma}=\frac{e E_0 N_u \lambda_u}{\gamma m c^2}\Big\langle \sin{\big(k_s(z\cos{\theta}+x\sin{\theta}-ct)+\Phi\big)} \nonumber\\&&\times (\beta_{x}\cos{\theta}-\beta_{z}\sin{\theta})\Big\rangle_z.
\end{eqnarray}
Simplify the item between the brackets with the help of Eq.~(11), we get
\begin{widetext}
\begin{eqnarray}
&&\Big\langle \sin{\big(k_s(z\cos{\theta}+x\sin{\theta}-ct)+\Phi\big)} \cdot (\beta_{x}\cos{\theta}-\beta_{z}\sin{\theta})\Big\rangle_z\nonumber\\
&=&\Big\langle \sin{\big(-\zeta+a\sin{(2\zeta)}+d\cos{\zeta}\big)} \cdot \big(o+p\sin{\zeta}+q\cos{(2\zeta)}\big)\Big\rangle_{\zeta} \cdot \cos{\Phi},
\end{eqnarray}
where $\zeta=k_u z$ means the longitudinal phase, $a =\frac{\xi}{2}$, $d=\frac{k_s K\sin{\theta}}{k_u\gamma}$, $o=\big(\frac{1}{2\gamma^2}(1+\frac{K^2}{2})-1\big)\cdot\sin{\theta}$, $p=-\frac{K\cos{\theta}}{\gamma}$ and $q=-\frac{K^2}{4\gamma^2}\sin{\theta}$. Now the brackets denote the average of the $\zeta$ coordinate over $2\pi$: $\langle\cdots\rangle_{\zeta}=\frac{1}{2\pi}\int_0^{2\pi} (\cdots) d\zeta$. Then the energy change is 
\begin{eqnarray}
\Delta\delta&=&\frac{e E_0 N_u \lambda_u}{\gamma m c^2}\Big\langle \sin{\big(-\zeta+a\sin{(2\zeta)}+d\cos{\zeta}\big)} \cdot \big(o+p\sin{\zeta}+q\cos{(2\zeta)}\big)\Big\rangle_{\zeta}\cdot \cos{\Phi}=\Gamma\cos{\Phi}\nonumber\\
&=&\Gamma \cos{(k_s (z_0\cos{\theta}+x_0\sin{\theta}))},
\end{eqnarray}
\end{widetext}
where $\Gamma=\frac{e E_0  N_u \lambda_u}{\gamma m c^2}\Big\langle \sin{\big(-\zeta+a\sin{(2\zeta)}+d\cos{\zeta}\big)} \cdot \big(o+p\sin{\zeta}+q\cos{(2\zeta)}\big)\Big\rangle_{\zeta}$. 

Now we define the instantaneous change of electron horizontal divergence due to the interaction with the laser electromagnetic field in the undulator:
\begin{eqnarray}
\frac{\mathrm{d}x^{\prime}}{\mathrm{d}z}=\frac{e}{\gamma m c^2}E_{x_l}(\cos{\theta}-\beta_{z}),
\end{eqnarray}
where $x^{\prime}=\frac{dx}{dz}$. Similar with the previous derivation procedure of energy modulation, one obtains the angular modulation expression:
\begin{eqnarray}
\Delta x^{\prime}=\Pi\cos{\Phi}=\Pi \cos{(k_s (z_0\cos{\theta}+x_0\sin{\theta}))},
\end{eqnarray}
where $\Pi=\frac{e E_0  N_u \lambda_u}{\gamma m c^2}\Big\langle \sin{\big(-\zeta+a\sin{(2\zeta)}+d\cos{\zeta}\big)} \cdot \big(u+v\cos{(2\zeta)}\big)\Big\rangle_{\zeta}$, $u=\frac{1}{2\gamma^2}(1+\frac{K^2}{2}-\gamma^2\theta^2)$ and $v=-\frac{K^2}{4\gamma^2}$.

With respect to the energy modulation $\Gamma$ and angular modulation amplitude $\Pi$, through tremendous numerical verifications, we get 
\begin{eqnarray}
\Pi=\Gamma\cdot \sin{\theta}.
\end{eqnarray}
With a small incident angle, Eq.~(23) and Eq.~(25) can be simplified to
\begin{subequations}
\begin{eqnarray}
\Delta\delta&=&\Gamma\cos{(k_s z_0+k_s x_0\theta)},\\
\Delta x^{\prime}&=&\Gamma \theta\cos{(k_s z_0+k_s x_0\theta)}.
\end{eqnarray}
\end{subequations}
This result is also in agreement with the Panofsky-Wentzel theorem $\frac{\partial}{\partial x_0} (\Delta\delta) = \frac{\partial \Delta x^{\prime}}{\partial z_0}$.

The above theoretical analyse shows that the obliquely incident laser and electron beam interaction in the undulator would not only imprint energy modulation but also angular modulation on the electron beam phase space, which extends the applications of the laser-electron interaction from two-dimensional to three-dimensional manipulations.

\subsection{Comparison of modulation amplitudes}
\begin{figure}
    \includegraphics[width=1\linewidth]{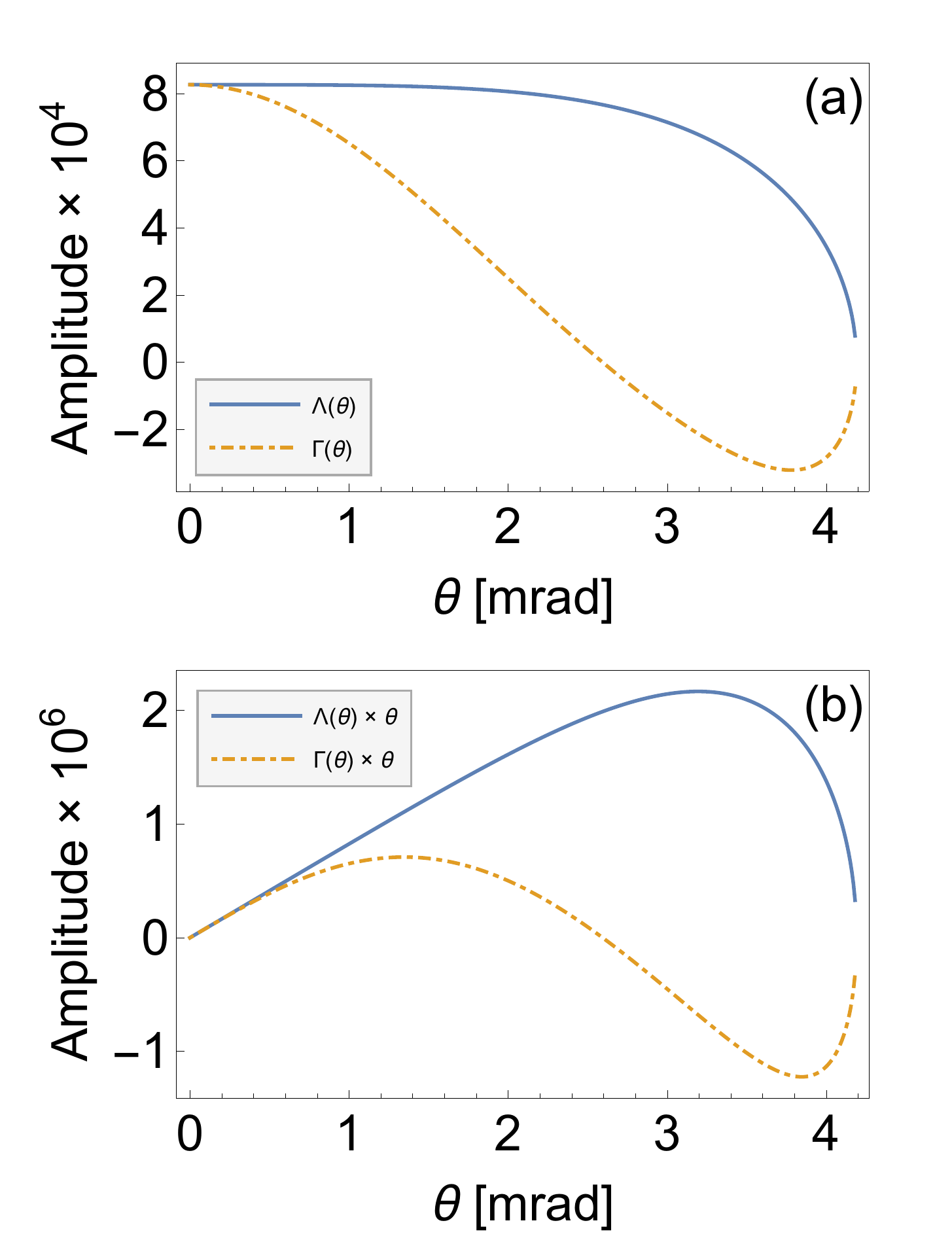}
    \caption{Laser-induced energy (a) and angular (b) modulation amplitudes as functions of the laser incident angle. $\Lambda(\theta)$ and  $\Lambda(\theta)\times \theta$ refer to the energy and angular modulation amplitudes of laser incident in the $z - y$ plane, respectively. $\Gamma(\theta)$ and $\Gamma(\theta)\times \theta$ refer to the energy and angular modulation amplitudes of laser incident in the $z - x$ plane.}
    \label{fig2}
\end{figure}

In the above discussion, two cases are taken into consideration: the laser incident in the $z - y$ plane and in the $z - x$ plane. Equation (13) and Eq.~(23) show that the energy modulation amplitudes $\Lambda$ and $\Gamma$ have different expressions. So do the angular modulation amplitudes $\Lambda\theta$ and $\Gamma \theta$. Under resonant condition (see Eq.~(5)), changes in the incident angle of the seed laser will drive changes in the undulator parameter, which will further change the laser-induced modulation amplitude. Figure~\ref{fig2} illustrates the energy and angular modulation amplitudes as functions of the laser incident angle 
based on the parameters in Table~\ref{tab1}.
\begin{table}
\caption{\label{tab1}The nominal parameters of electron beam, laser and undulator.}
\begin{ruledtabular}
\begin{tabular}{ll}
Beam energy & 1.5 GeV \\
Relative energy spread & $0.01\%$\\
Geometric horizontal or vertical emittance & 0.1 nm rad\\
Laser electric field & 5 GV/m \\
Laser wavelength & 265 nm\\
Undulator period & 3 cm \\
Number of Undulator periods               & 4\\
\end{tabular}
\end{ruledtabular}
\end{table}

When the incident angle equals to zero, the laser pointing direction coincides with the propagation direction of the electron beam. This corresponds to the normally incident pattern with the energy modulation amplitude of $8.26\times 10^{-4}$. Curves of $\Lambda(\theta)$ and $\Gamma(\theta)$ all start with this point. One can see from Fig.~\ref{fig2}(a) that, $\Lambda(\theta)$ decreases as the angle increases and the downtrend gets fast as the angle increases. $\Gamma(\theta)$ firstly drops to zero and keeps the decreasing tendency and finally increases back to nearly zero. Equation (5) shows that under the resonant condition, the undulator parameter decreases as the incident angle increases. When the undulator parameter is reduced to zero, the energy transfer between laser and electron beam equals to zero. This explains why these two functions all eventually approaches zero when the incident angle becomes large. As for the zero-crossing behavior of $\Gamma(\theta)$, this is because, for the laser incident in the $z - x$ plane case, components of the laser electric field in the $x$ and $z$ directions induce opposite energy modulations, as shown in Eq.~(20). As the angle increases, the ratio of the energy modulations introduced by these two components changes, which causes the modulation amplitude transfer from positive to negative. 

The angular modulation amplitudes equal to the energy modulation amplitudes multiplied by the incident angle. Compared to the energy modulation, the curves of the angular modulation amplitude exhibit different trends, as can be seen from Fig.~\ref{fig3}(b). These two curves all start from zero and end up with zero. For the case of laser incident in the $z$ - $y$ plane, the angular modulation amplitude firstly rises and then decreases to nearly zero. It is apparent that there is an optimal value for the angular modulation amplitude. For the laser incident in the $z$ - $x$ plane case, the angular modulation amplitude firstly increases, then maintains the downtrend and passes through zero and finally increases back to zero. Two peaks of the angular modulation amplitude appear in this case. These analyses reveal the complexity of laser incident in the $z$ - $x$ plane case over the case of laser incident in the $z$ - $y$ plane.   

\subsection{Simulations}
\begin{figure}
    \includegraphics[width=1\linewidth]{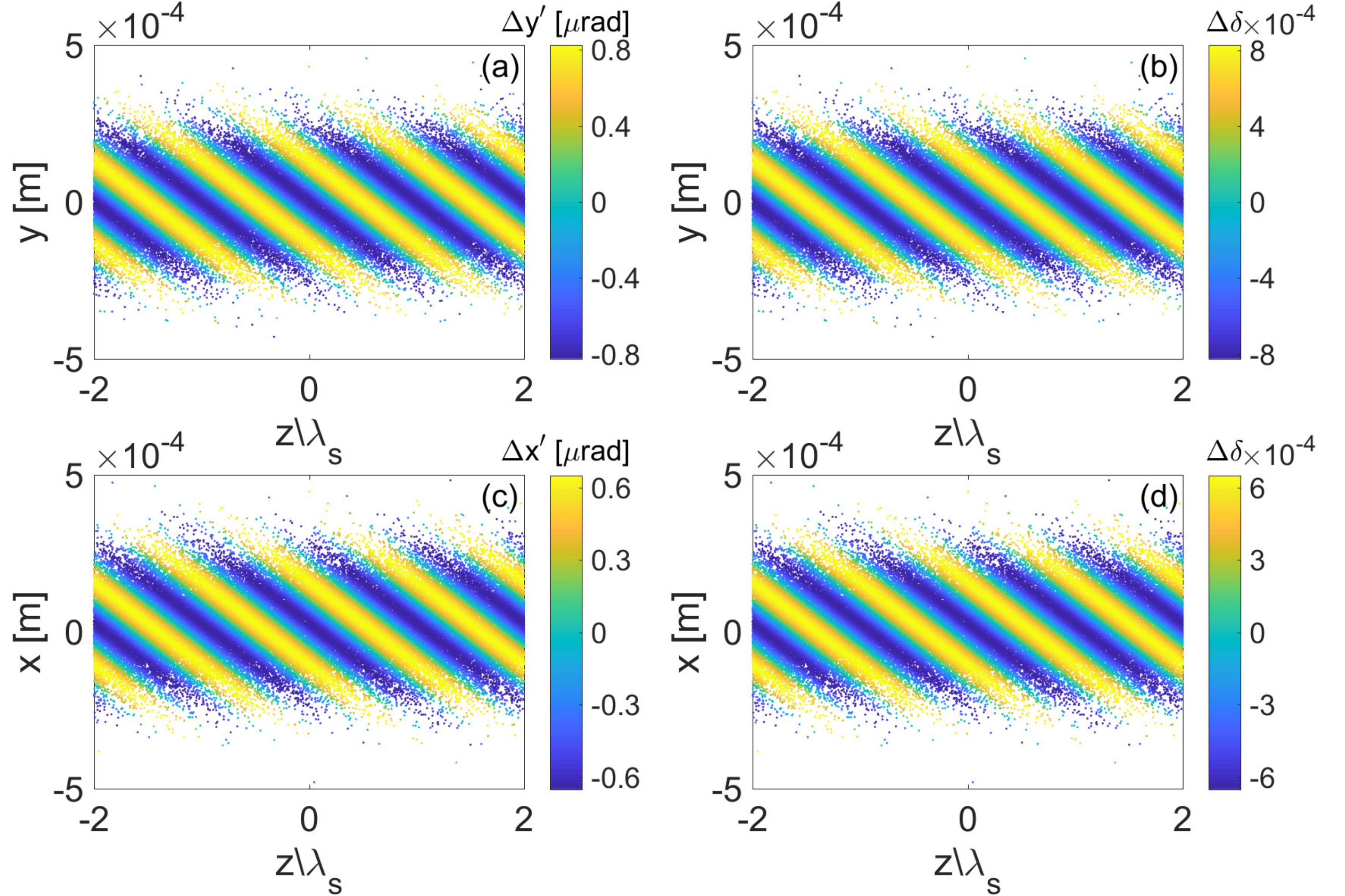}
    \caption{Distributions of the induced divergences and energy deviations after the obliquely incident laser-electron interaction with a 3D simulation program. (a) and (b) refer to the laser incident in the $z-y$ plane case; (c) and (d) refer to the laser incident in the $z-x$ plane. For these two cases, the incident angles of the seed laser are both $1$ mrad.}
    \label{fig3}
\end{figure}

To demonstrate the previous theoretical results and show the physical mechanism of the angular modulation, three-dimensional (3D) simulations are performed in this section. On the basis of Eqs.~(\ref{eq1}), (5 - 9), (14) and (18 - 20), two 3D and time-dependant codes are developed to numerically study the obliquely incident laser-electron interaction in an undulator. One for the laser incident in the $z - y$ plane case, the other for the laser incident in the $z - x$ plane case. In our simulation programs, the Lorentz force acting on the electrons comes from the laser electromagnetic field and the undulator magnetic field:
\begin{eqnarray}
\frac{\mathrm{d}x^{\prime}}{\mathrm{d}z}=\frac{e}{\gamma m c^2}\Big( E_x  (1-\beta_z \cos\theta )-  v_z  B_y\Big)
\end{eqnarray}
for the first case, and
\begin{eqnarray}
\frac{\mathrm{d}x^{\prime}}{\mathrm{d}z}=\frac{e}{\gamma m c^2}\Big( E_x  (\cos\theta-\beta_z )-  v_z  B_y\Big)
\end{eqnarray}
for the second case. For each case, $\beta_{z}=\sqrt{1-\frac{1}{\gamma^2}-\beta_{x}^2}$ is used to calculate the longitudinal velocity of the electron beam. These formulas are directly used to simulate the electron motion in the undulator instead of Eq.~(2) and (3). This method may introduce some nonlinear terms with reference to the theoretical derivation. However, it is closer to the real situation. The entire undulator line is divided into adequate integration steps for an accurate solution. The numerical simulations illustrate the distributions of the laser-induced vertical divergence and energy deviation in Fig.~\ref{fig3}.

The simulations also use parameters in Table~\ref{tab1} with incident angle of $\theta=1$ mrad. We denote $\beta_x$ and $\beta_y$ as the beta functions at the entrance of the modulator. For the first case, the root-mean-square (RMS) beam size $\sigma_y$ would be $1\times 10^{-4}$ with $\beta_y=100$. For the second case, $\beta_x=100$, then $\sigma_x$ is $1\times 10^{-4}$. From Eq.~(13) and (23) , one can calculate the energy modulation amplitudes of $8.3\times10^{-4}$ for the first case and of $6.5\times10^{-4}$ for the second case (see Fig.~\ref{fig2}). These two theoretically predicted values are confirmed by our simulations, as can be seen from Fig.~3(b) and 3(d). Meanwhile, Fig.~\ref{fig3} shows that the divergence and energy are both periodically modulated and the longitudinal distance between the adjacent modulation peaks is exactly one wavelength of the seed laser. The coefficient ratio of the induced angle modulation and the energy modulation equals to the incident angle $\theta$. These results are consistent with Eq.~(17) and Eq.~(27).

\begin{figure}
    \includegraphics[width=1\linewidth]{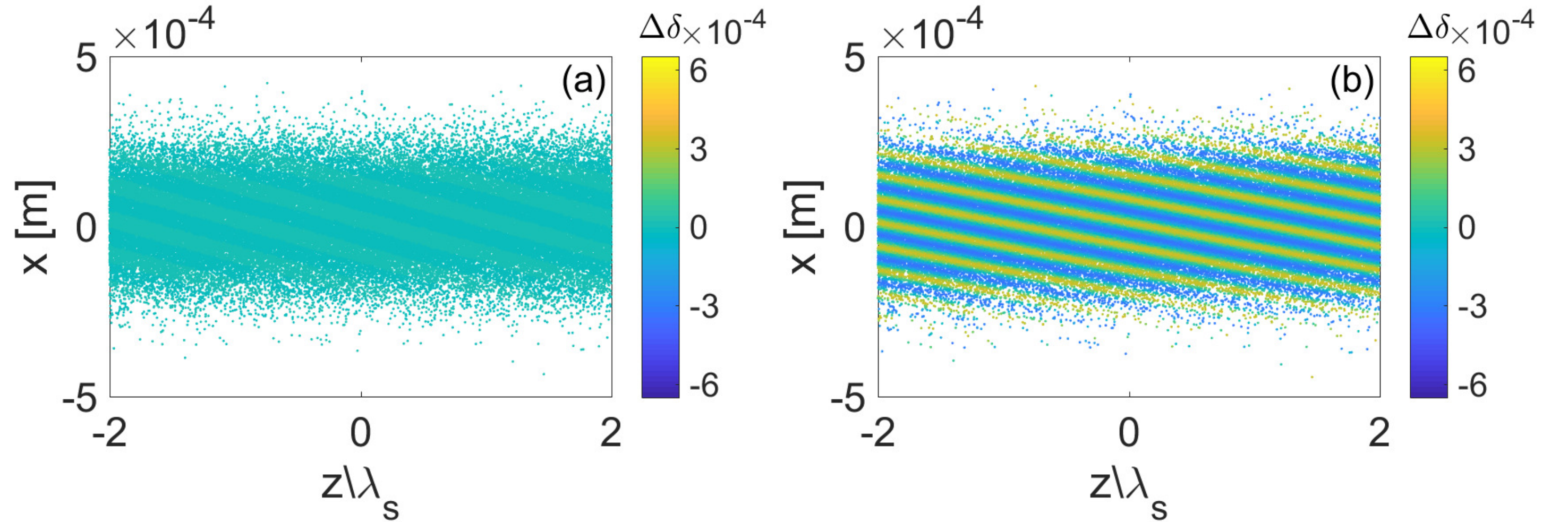}
    \caption{For the case of laser incident in the $z$ - $x$ plane, distributions of the induced energy deviations after the obliquely incident laser-electron interaction with a 3D simulation program. The incident angles of the seed laser are $2.6$ mrad (a) and $3.8$ mrad (b), respectively.}
    \label{fig4}
\end{figure}

Further simulations are performed to demonstrate the evolution of the energy modulation amplitude with the laser incident angle for the case of laser incident in the $z$ - $x$ plane. The simulation results are shown in Fig.~\ref{fig4} 
with laser incident angle of $2.6$ mrad (Fig.~\ref{fig4}(a)) and of $3.8$ mrad (Fig.~\ref{fig4}(b)). From Fig.~\ref{fig2}, one can see that $2.6$ mrad and $3.8$ mrad are close to the zero-crossing point and the minimum point, respectively. With the help of Eq.~(23), we get the energy modulation amplitudes at these two points: $-2.5\times 10^{-6}$ and $-3.2\times 10^{-4}$, respectively. The negative sign represents the opposite phase of the laser-induced energy modulation. Our simulations verify these two values. With the same scaled colorbar, we can directly see the relatively changes in modulation amplitude (Fig.~\ref{fig4}) in comparison with that of $1$ mrad incident-angle case (Fig.~\ref{fig3}(d)). We also observe the phase change in the simulations. For the electron in a fix position, like $(x,z)=(0,-2\lambda_s)$, the laser-induced energy modulation amplitudes are negative with incident angle of $1$ mrad (Fig.~\ref{fig3}(d)) and positive with incident angle of $3.8$ mrad (Fig.~\ref{fig4}(b)), respectively. 

Whether for the case of laser incident in the $z$ - $y$ plane or for the case of laser incident in the $z$ - $x$ plane, our simulation results are consistent with the theoretical predictions. And the differences between the two cases are revealed.

\section{Angular drift effects}\label{sec3}
The above study shows that the obliquely incident laser and electron beam interaction in an undulator would induce multi-dimensional modulations on the electron beam phase space, which is quite different from the normally incident pattern. It means that the angular drift of the seed laser would have a direct impact on mechanisms that rely on laser-electron modulation techniques, such as HGHG and EEHG. In this section, we firstly analyze the influences of the laser incident angle on the bunching factors of HGHG and EEHG, respectively. Then we discuss the influences of the laser angular drift occurring in HGHG or EEHG on the emittance of the electron beam. For simplicity, we only consider the case of laser incident in the $z - y$ plane.

\subsection{Bunching factor tuning of HGHG}
Bunching factor is a physical quantity to quantify the density modulation of the electron beam which contains high harmonic components. In the following, we'll derive expression of the bunching factor of HGHG under the condition of obliquely incident laser modulation. 
To simplify the derivation, we use the normalized energy deviation $P=\delta / \sigma_\delta$ and dimensionless vertical position of a particle $Y=y/\sigma_y$ as main variables, where $\sigma_\delta$ is the RMS of the energy deviation $\delta$, $\sigma_{y}=\sqrt{\varepsilon_y\beta_y}$ is the intrinsic vertical RMS beam size. $\varepsilon_y$ and $\beta_y$ are respectively the beam vertical emittance the beta function at the entrance of the modulator. Normally, the definition of bunching factor can be written as:
\begin{equation}
b=\frac{1}{N} \iint \mathrm{d}Y \mathrm{d}P f(Y,P) \langle e^{-i \mathtt{a} \zeta}\rangle_{\zeta},
\end{equation}
where $N$ is the total number of electrons, $\mathtt{a}$ is a number, $f(Y,P)$ is the function of beam distribution:
\begin{equation}
f(Y,P)=\frac{N}{2\pi}e^{-\frac{Y^2}{2}}e^{-\frac{P^2}{2}}.
\end{equation}

With an incident angle $\theta$, the manipulation expression of the electron beam in the modulator is different from the normally incident pattern, and can be written as
\begin{equation}
P_1=P+A \sin(\zeta + k_s y \theta),
\end{equation}
where $A=\Lambda/\sigma_\delta$. Chicane converts the energy modulation into density modulation
\begin{equation}
\zeta_1= \zeta +B P_1,
\end{equation}
where $B=R_{56} k_s \frac{\sigma_{\gamma}}{\gamma}$ is the dimensionless dispersion strength of the chicane. $R_{56}$ is the momentum compaction of the chicane. After the electron beam traversing the dispersive section, a strong beam density can be obtained. With Eqs.~(30 - 33), we get the expression of bunching factor
\begin{eqnarray}
  b_n=e^{-\frac{(nk_s\theta)^2\varepsilon_y\beta_y}{2}}\cdot e^{-\frac{(nB)^2}{2}}  \cdot J_n(-n A B),
\end{eqnarray}
where $n$ represents the harmonic order. Compared with normally incident pattern, the oblique incidence of the seed laser brings two major effects: the change in modulation amplitude and the wavefront tilt of the bunching. The modulation amplitude change would affect the value of the Bessel function. The wavefront tilt of the bunching induces the first exponential term in Eq.~(34). The exponential term is related to the laser incident angle and beta function (or initial beam size).

\begin{figure}
    \includegraphics[width=1\linewidth]{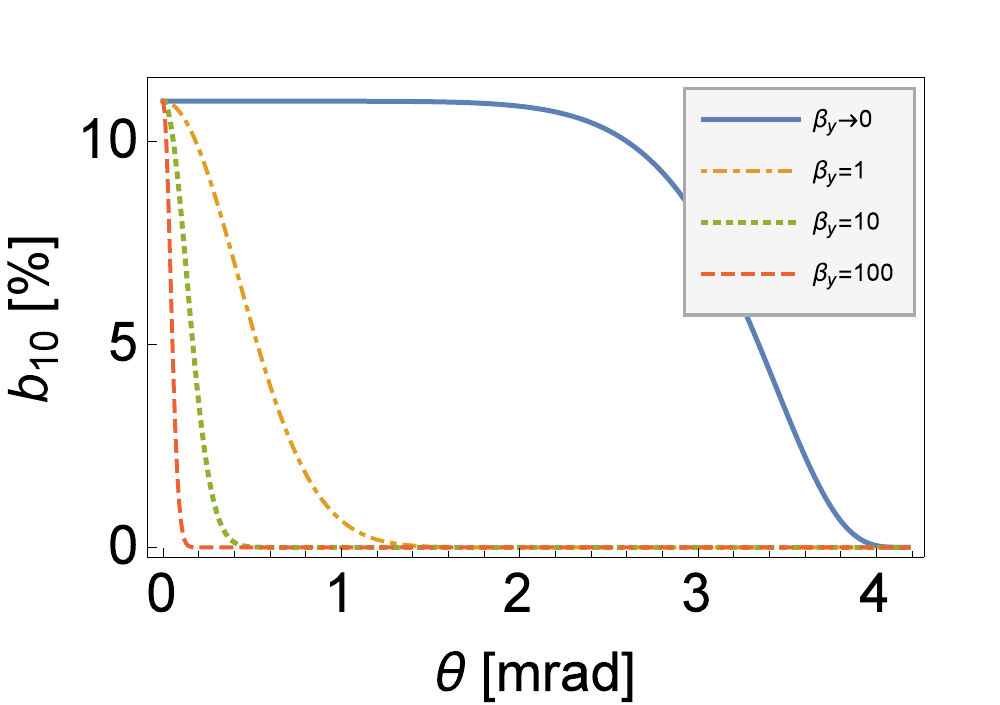}
    \caption{In the HGHG mechanism, the bunching factor at 10-th harmonic $b_{10}$ as a function of the laser incident angle $\theta$ with different beta functions.}
    \label{fig5}
\end{figure}

Figure~\ref{fig5}
illustrates the bunching factor at 10-th harmonic as a function of the laser incident angle with different beta functions in HGHG mechanism. The study uses the parameters in Table~\ref{tab1}. The dispersion strength of the chicane is $60.0$ $\mu$m and it does not change with incident angles. Then the bunching factor under normally incident pattern is $11.0\%$. One can see from Fig.~\ref{fig5} that the bunching factor decreases as the incident angle increases regardless of the beta function. When $\beta_y\rightarrow 0$, the vertical beam size is very small, then the bunching tilt effect could be ignored. The bunching factor would only affect by the modulation amplitude change due to the angular drift. As the beta function increases, the vertical beam size of the electrons becomes larger, and the bunching tilt effect plays a dominant role in determining the bunching factor. It causes the downtrend of the bunching factor gets faster as the beta function increases.

\subsection{Bunching factor tuning of EEHG}
EEHG contains two modulator magnets and two chicanes. The laser angular drift of the seed laser may occur in any of these two modulators. For the cases where two seed lasers are both normally incident in two modulators, which refers to the nominal EEHG, the bunching factor at the $\mathtt{a}$-th harmonic is given by~\cite{stupakov2009using}:
\begin{equation}
  b_{\mathtt{a}}=\sum_{n,m}e^{-\frac{C^2}{2}}\cdot J_n(-CA_1)\cdot J_m(-\mathtt{a} A_2 B_2),
\end{equation}
where $\mathtt{a}=n+m \mathcal{K}$, $C=n B_1+\mathtt{a} B_2$. $n$ and $m$ are integers of either sign. $\mathcal{K}=k_{s_2}/k_{s_1}$ is the ratio of wave number of the second seed laser to the first. $A_1$ and $A_2$ are the energy-modulation amplitudes of the first and second modulator, respectively. $B_1$ and $B_2$ are the dimensionless dispersion strengths of the first and second chicane, respectively.

When the laser angular drift only occurs in the first modulator, the bunching factor at the $\mathtt{a}$-th harmonic is derived as:
\begin{equation}
 b_{\mathtt{a}}=\sum_{n,m}e^{-\frac{(n k_{s_1}\theta)^2\varepsilon_y\beta_y}{2}}\cdot e^{-\frac{C^2}{2}}\cdot J_n(-CA_1)\cdot J_m(-\mathtt{a} A_2 B_2),
\end{equation}
where an exponential term is added in the expression pertaining to Eq.~(35). This term contains integer n, wave number of the first seed laser $k_{s1}$, incident angle $\theta$, vertical beam emittance $\varepsilon_y$ and beta function $\beta_y$.

When the laser angular drift only occurs in the second modulator, the bunching factor at the $\mathtt{a}$-th harmonic reads:
\begin{equation}
  b_{\mathtt{a}}=\sum_{n,m}e^{-\frac{(m k_{s_2}\theta)^2\varepsilon_y\beta_y}{2}}\cdot e^{-\frac{C^2}{2}}\cdot J_n(-CA_1)\cdot J_m(-\mathtt{a} A_2 B_2).
\end{equation}
An exponential term also occurs in the expression pertaining to Eq.~(35). This term contains integer m, wave number of the second seed laser $k_{s2}$, incident angle $\theta$, vertical beam emittance $\varepsilon_y$ and beta function $\beta_y$.
Previous study shows that n and m should be of opposing signs to ensure the dispersion $B_1$ and $B_2$ have the same sign and $n=\pm 1$ to have the maximum harmonic bunching~\cite{xiang2009echo}. Hence one can see from Eq.~(36) and (37) that when  the two seed lasers possess the same wavelength, the laser angular drift in the second modulator has a much enormous impact on the high harmonic bunching than that in the first modulator. 

\begin{figure}
    \includegraphics[width=1\linewidth]{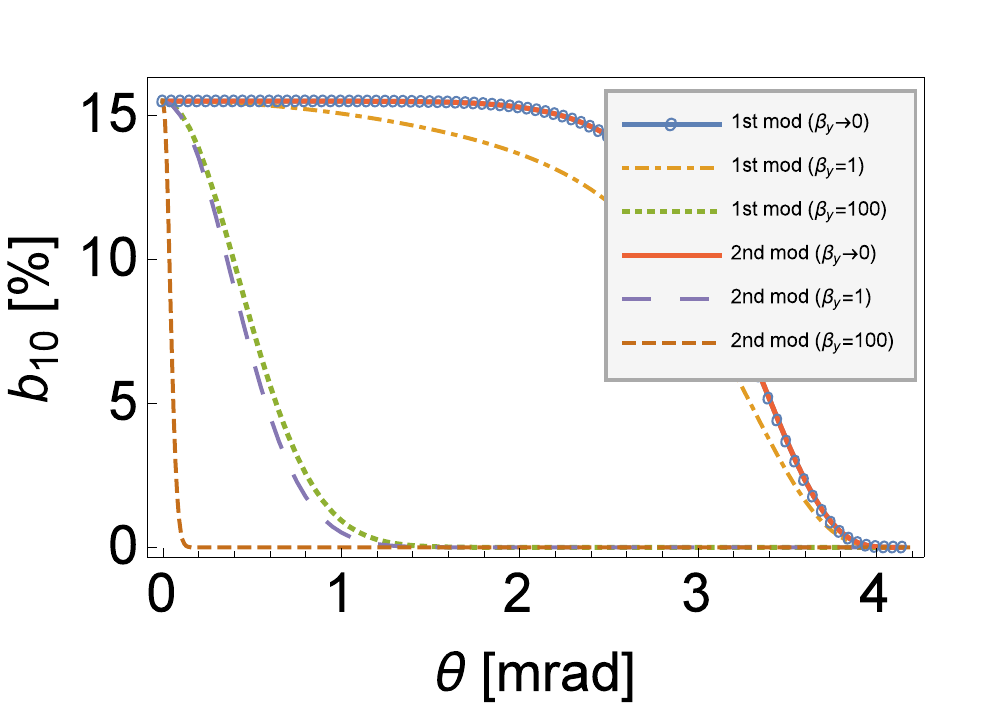}
    \caption{In the EEHG mechanism, the bunching factor at 10-th harmonic $b_{10}$ as a function of the laser incident angle $\theta$ with different beta functions. "1st mod" refers to the laser angular drift occurring only in the first modulator. So does "2nd mod".}
    \label{fig6}
\end{figure}

Figure~\ref{fig6}  
displays the bunching factor at 10-th harmonic as a function of the laser incident angle with different beta functions in EEHG mechanism. The study uses the parameters in Table~\ref{tab1} except for the laser electric field. The electric fields of both seed lasers are $2.5$ GV/m. The dispersion strengths of the first and second chicane are respectively $1.50$ mm and $0.13$ mm and they do not change with incident angle. The bunching factor under normally incident pattern is $15.5\%$. Figure~\ref{fig6} tells that regardless of the beta functions and of which modulator existing laser angular drift, the bunching factor decreases as the incident angle increases. When we ignore the bunching tilt effect ($\beta_y \rightarrow 0$), the bunching factor is only affected by the modulation amplitude variation. For either modulator with laser angular drift, the curve of the bunching factor is nearly overlapping. And for either modulator with laser angular drift, the bunching factor with large beta function (beam size) is more sensitive to the laser incident angle with reference to that with small one. Moreover, for the electron beam with a certain beam size, the bunching degradation caused by the laser angular drift in the second modulator is much more serious than that caused by the first one. 

\subsection{Emittance growth}
In the mechanism of HGHG or EEHG, the obliquely incident laser would introduce not only energy modulation but also angular modulation. It is apparent that angular modulation would cause an increase in the emittance of the electron beam. In this section, we analyze the effect of laser angular drift on the electron beam emittance.

\begin{figure}
    \includegraphics[width=1\linewidth]{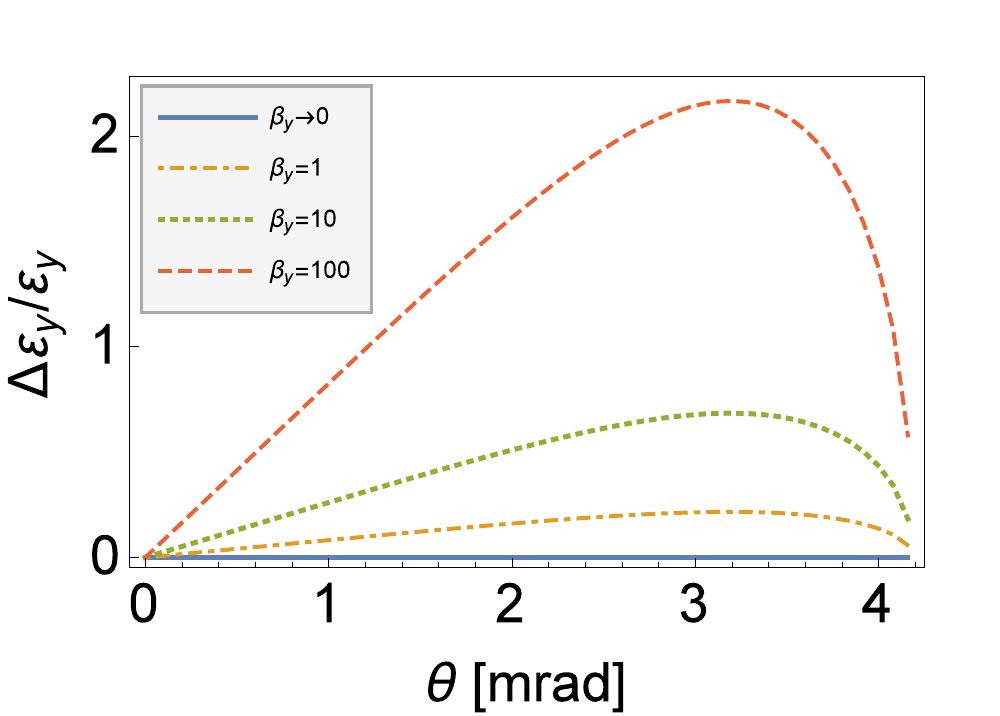}
    \caption{Normalized emittance growth as a function of the laser incident angle $\theta$ with different beta functions.}
    \label{fig7}
\end{figure}

It is assumed that the Twiss parameter $\alpha$ at the entrance of the undulator equals to zeros, then the vertical RMS divergence of the electron beam is $\sigma_{y^{\prime}}=\sqrt{\varepsilon_y/\beta_y}$. With the help of Eq.~(17b), we get the expression of the normalized emittance growth 
\begin{equation}
\frac{\Delta \varepsilon_y}{\varepsilon_y}\propto \frac{\Delta y^{\prime}}{\sigma_{y^{\prime}}} \approx \Lambda\theta \sqrt{\frac{\beta_y}{\varepsilon_y}},
\end{equation}
where we ignore the cosine term to predict the maximum change of emittance. Figure~\ref{fig7}
illustrates the normalized emittance growth as a function of the laser incident angle with different beta functions. We also use the parameters in Table~\ref{tab1}. With $\beta_y \rightarrow 0$, the beam vertical divergence is extremely large, then the induced angular modulation would not have any impact on the beam emittance. For the other cases, the tendencies of the curves all firstly rise and then decrease to nearly zero, which are consistent with the tendency of the angle modulation curve in Fig.~\ref{fig2}(b). For the case with a large $\beta_y$, the initial electron divergence is small. Even with the same laser-induced angular modulation as in other cases, the modulation has a relatively large contribution to the beam emittance.

In HGHG or EEHG, the bunching factor is very sensitive to the laser incident angle. The laser incident angle should be very small to maintain a relatively large bunching factor. Hereby one can see from Fig.~\ref{fig7}, the induced emittance growth would be very small when the incident angle is close to zero.

\section{Summary and discussions}\label{sec4}
Frequency up-conversion schemes have the ability to significantly improve the temporal coherence and inhibit the fluctuations of high-gain FELs. External seed lasers are typically used to manipulate the electron beam phase space in these schemes. Therefore, the performances of these schemes are largely affected by the property of the external seed lasers, such as angular drift. 

Laser oblique incidence is the obvious manifestations of the angular drift. In this work, we systematically study the obliquely incident laser-electron interaction in the undulator. Theoretical analysis and numerical simulations demonstrate that the interaction would imprint not only energy modulation but also angular modulation on the electron beam phase space. Further studies show that laser angular drift results in reductions in the energy modulation amplitude and bunching factors of HGHG and EEHG. It is also found that in HGHG and EEHG, the bunching factors are more sensitive to laser incident angles for the electrons with a relatively large beam size (beta functions). And the angular drift which occurs in the second modulator of EEHG has a greater impact on the bunching factor than that in the first modulator.  Moreover, the laser-induced angular modulation would slightly increase the emittance of the electron beam.

Generally, angular drift occurs in HGHG or EEHG mechanism passively and inevitably.  
Despite the fact that angular drift interferes with the performance of HGHG and EEHG, it still can be seen as a multi-dimensional modulation technique. Generating multi-dimensional modulation with one component is of great application prospect in the field of laser-electron manipulation, therefore the discussed technique may bring fresh air to this area. Several mechanisms~\cite{feng2014three,feng2015generating,wang2019angular} have been proposed based on this technique. It should be emphasized that the laser-induced angular modulation would increase the divergence then further spoil the emittance of the electron beam, which should be carefully considered in every specific scheme. Another key drawback of this technique is that the laser requires a large transverse size, which corresponds to a strong laser power, to fully cover the electron beam. This is detrimental to mechanisms that require high laser power. 

Further investigations on these topics are still ongoing.

\begin{acknowledgments}
The authors would like to thank Alex Chao (SLAC) and Juhao Wu (SLAC) for helpful discussions and useful comments related to this work. This Work is supported by the National Nature Science Foundation of China (Grant No. 11605277) and National Key Research and Development Program of China (No. 2016YFA0401901).
\end{acknowledgments}


\bibliography{mybibfile}

\end{document}